\documentclass[prd,preprint,showpacs,preprintnumbers,amsmath,amssymb,eqsecnum]{revtex4}

\usepackage{graphicx}
\usepackage{dcolumn}
\usepackage{bm}
\usepackage{subfigure}
\usepackage{color}

\begin{document}

\preprint{RUP 11-2, YITP-11-75}

\title{
Collision of an object in the transition from adiabatic inspiral to plunge 
around a Kerr black hole
}
\author{$^{1}$Tomohiro Harada}%
 \email{harada@rikkyo.ac.jp}
\author{$^{2}$Masashi Kimura}
\email{mkimura@yukawa.kyoto-u.ac.jp}
\affiliation{%
$^{1}$Department of Physics, Rikkyo University, Toshima, Tokyo 171-8501, Japan\\
$^{2}$Yukawa Institute for Theoretical Physics, Kyoto 606-8502, Japan
}%
\date{\today}

\begin{abstract}
An inspiralling object of mass $\mu$ around a Kerr black hole
of mass $M (\gg \mu)$ 
experiences a continuous transition near the innermost 
stable circular orbit from adiabatic inspiral to 
plunge into the horizon as gravitational radiation 
extracts its energy and angular momentum.  
We investigate the collision of 
such an object with a generic counterpart
around a Kerr black hole.
We find that the angular momentum of the object is 
fine-tuned through gravitational radiation and 
that the high-velocity collision of the object 
with a generic counterpart naturally occurs around a nearly maximally 
rotating black hole. We also find that the 
centre-of-mass energy can be far beyond the Planck energy
for dark matter particles colliding around a stellar mass 
black hole and as high as $10^{58}$ erg for 
stellar mass 
compact objects colliding around a supermassive black hole,
where the present transition formalism is well justified.
Therefore, rapidly rotating black holes can 
 accelerate objects inspiralling around them to 
energy high enough to be of great physical interest. 
\end{abstract}

\maketitle


\newpage

\section{Introduction}
Ba\~nados et al.~\cite{Banados:2009pr} 
discovered that the centre-of-mass (CM) energy can be arbitrarily high
for the collision of two geodesic particles moving on the 
equatorial plane around a nearly maximally rotating Kerr black hole.
The angular momentum of either of the particles must be 
artificially fine-tuned for such a striking event.
This phenomenon is seen not only for Kerr black holes
but also for Kerr-Newman black holes~\cite{Wei:2010vca},
exotic black holes~\cite{Wei:2010gq}, and 
naked singularities~\cite{Patil:2011ya,Patil:2011uf}. 
The analysis is extended to the collision of 
particles in nonequatorial motion 
for Kerr black holes~\cite{Harada:2011xz}, Kerr-Newman black 
holes~\cite{Liu:2011wv}, and accelerating and rotating black holes~\cite{Yao:2011ai}.
The general explanation of this phenomenon is proposed 
in Ref.~\cite{Zaslavskii:2011dz}.
This phenomenon is studied in the astrophysical 
contexts of dark matter particle 
annihilation~\cite{Banados:2009pr,Banados:2010kn,Williams:2011uz},  
extreme mass-ratio inspirals (EMRIs), and accretion 
disks~\cite{Harada:2011xz,Harada:2010yv}. 

It is argued that 
the effects of gravitational radiation would constrain
the maximum CM energy 
because the particle with the fine-tuned angular momentum 
can reach the horizon after it orbits around 
the black hole infinitely many times in infinitely 
long proper time~\cite{Berti:2009bk,Jacobson:2009zg}. 
On the other hand, the effects of conservative 
self-force bound the CM energy from 
above in the analogous system of spherical charged 
shells~\cite{Kimura:2010qy}.
It is also argued~\cite{Berti:2009bk,Jacobson:2009zg} 
that the CM energy cannot be extremely high
because the nondimensional spin of astrophysical black holes 
is bounded by Thorne's limit 0.998~\cite{Thorne:1974ve}.
However, it is not clear whether there is a universal bound strictly 
less than unity
on the black hole spin, as Thorne's limit is thought to be dependent 
on the accretion flow models~\cite{Abramovitz:1980,Popham:1998,Sadowski:2011}.

As for the fine-tuning problem, the present 
authors~\cite{Harada:2010yv} proposed 
a scenario where the fine-tuning is realised in 
EMRIs.
Since the ratio of the gravitational radiation time scale $t_{\rm GW}$
to the orbital period $t_{\rm orb}$ is given by $t_{\rm GW}/t_{\rm orb}\sim \eta^{-1}$, where $\eta=\mu/M$ is the mass ratio, the inspiral 
through gravitational radiation will be regarded as 
adiabatic if $\eta\ll 1$. Noting the circularisation of 
the orbits in the post-Newtonian regime~\cite{Peters:1963ux}, 
we can assume that 
an inspiralling compact object adiabatically takes a 
circular orbit which is closer to the black hole as the object loses 
its energy and angular momentum through gravitational radiation.
Once the compact object reaches the radius of the innermost stable circular orbit (ISCO), it begins to plunge
into the black hole in the dynamical time scale.
Thus, the compact object will eventually have 
the energy and angular momentum of the particle orbiting the ISCO.
In the maximal rotation limit of the black hole, the fine-tuning 
of the angular momentum is realised for the ISCO particle. 
However, this scenario should be reconsidered carefully, when 
we take radiation reaction into account seriously. Although 
radiation reaction drives 
the inspiralling object inwardly, 
it also gives the object an inward radial velocity at 
the ISCO radius, implying that 
the energy and angular momentum of the 
compact object are no longer those of the ISCO particle.
In such a situation, the formalism proposed by 
Ori and Thorne~\cite{Ori:2000zn} 
to describe the transition from 
adiabatic inspiral to plunge into a Kerr black hole
is quite useful. This formalism is 
extended to restore the consistency 
with the normalisation of the four-velocity
by Kesden~\cite{Kesden:2011ma}.

In the present paper, we apply the Ori-Thorne-Kesden formalism
for nearly maximally 
rotating black holes and estimate the CM energy for the collision 
of an object in the transition with a generic counterpart object. 
We find that the scenario proposed by the present authors~\cite{Harada:2010yv} is justified: the fine-tuning of the angular momentum is realised by the object in the transition from inspiral to plunge 
through gravitational radiation 
and the CM energy for the collision 
can be significantly high. Under the condition
for the Ori-Thorne-Kesden formalism to be justified, 
the CM energy can be much higher than the Planck energy for
dark matter particles colliding around a stellar mass black hole and 
can be as high as $10^{58}$ erg for compact objects colliding around a
supermassive black hole.
As another application, based on the present framework, 
we discuss that radiation reaction gives subdominant 
contributions to the proposal that 
a nearly maximally rotating black hole may be overspun by plunging an 
object~\cite{Jacobson:2009kt,Barausse:2010ka}.

This paper is organised as follows. In Sec.~II, 
we introduce the CM energy of two colliding particles 
around a Kerr black hole and its near-horizon limit.
In Sec.~III, we briefly review the Ori-Thorne-Kesden formalism 
of the transition from adiabatic inspiral to plunge.
In Sec.~IV, we apply the Ori-Thorne-Kesden formalism to
nearly maximally rotating black holes.
In Sec.~V, based on the Ori-Thorne-Kesden formalism, 
we estimate the CM energy for the collision of a transition
object with a generic counterpart. 
Section~VI is devoted to the conclusion.
In the Appendix, we revisit how the energy and the angular momentum radiated during the transition affect 
the spin of the final black hole in the merger with an 
inspiralling object.
We use the units in which $c=G=1$ and the abstract index notation
of Wald~\cite{Wald:1984rg}.

\section{CM energy of particles colliding around a Kerr black hole}

The line element in the Kerr spacetime 
in the Boyer-Lindquist coordinates is given by~\cite{Kerr1963,Wald:1984rg}   
\begin{eqnarray}
ds^{2}&=&-\left(1-\frac{2Mr}{\rho^{2}}\right)dt^{2}
-\frac{4Mar\sin^{2}\theta}{\rho^{2}}d\phi dt
+\frac{\rho^{2}}{\Delta}dr^{2}+\rho^{2}d\theta^{2} \nonumber \\
&&+\left(r^{2}+a^{2}+\frac{2Mra^{2}\sin^{2}\theta}{\rho^{2}}\right)
\sin^{2}\theta d\phi^{2} ,
\end{eqnarray}
where $a$ and $M$ are the spin and mass parameters, respectively,
$\rho^{2}=r^{2}+a^{2}\cos^{2}\theta$ and $\Delta=r^{2}-2Mr+a^{2}$.
If $0\le a^{2}\le M^{2}$, 
$\Delta$ vanishes at $r=r_{\pm}=M\pm\sqrt{M^{2}-a^{2}}$, where 
$r=r_{+}$ and $r=r_{-}$ correspond to an event horizon and 
Cauchy horizon, respectively. Here, we denote $r_{+}=r_{H}$. 
The surface gravity of the Kerr black hole is given by
$
\kappa_{H}=\sqrt{M^{2}-a^{2}}/(r_{H}^{2}+a^{2}).
$
Thus, the black hole has a vanishing surface gravity and hence is 
extremal for the maximal rotation $a^{2}=M^{2}$, while 
it is subextremal for the nonmaximal rotation $a^{2}<M^{2}$. 
The angular velocity of the 
horizon is given by 
\begin{equation}
\Omega_{H}=\frac{a}{r_{H}^{2}+a^{2}}.
\end{equation}
We can assume $a\ge 0$ without loss of generality. 

Let particles 1 and 2 of rest masses $\mu_{1}$ and $\mu_{2}$ 
have four-momenta 
$p^{a}_{1}$ and $p^{a}_{2}$ at the same spacetime point, respectively.
The CM energy $E_{\rm cm}$ of the two particles is then defined by
\begin{equation}
E_{\rm cm}^{2}=-(p_{1}^{a}+p_{2}^{a})(p_{1a}+p_{2a})=\mu_{1}^{2}+\mu_{2}^{2}-2g^{ab}p_{1a}p_{2b}.
\end{equation}
The derivation of the expression for the CM energy of two 
general 
particles around a Kerr black hole is 
described in detail in the authors' previous 
papers~\cite{Harada:2010yv,Harada:2011xz}. We do not repeat it here but quote the formula for the particles 
moving on the equatorial plane, where
$\theta=\pi/2$ and the Carter constant identically vanishes.
Equation~(3.2) of Ref.~\cite{Harada:2011xz} then reduces to
\begin{equation}
E_{\rm cm}^{2}=\mu_{1}^{2}+\mu_{2}^{2}+\frac{2}{r^{2}}\left[\frac{{\cal P}_{1}{\cal P}_{2}-\sigma_{1r}\sqrt{{\cal R}_{1}}\sigma_{2r}\sqrt{{\cal R}_{2}}}{\Delta}-(L_{1} -a E_{1})(L_{2}-a E_{2})\right],
\label{eq:E_cm_explicit_equatorial}
\end{equation}
where $\sigma_{ir}=\mbox{sgn}(p_{i}^{r})$, $E_{i}=-p_{it}$, 
$L_{i}=p_{i\phi}$,  
\begin{eqnarray}
{\cal R}_{i}&=& {\cal R}_{i}(r)={\cal P}_{i}(r)^{2}-\Delta(r) [\mu_{i} ^{2}r^{2}+(L_{i}-aE_{i})^{2}], \\
{\cal P}_{i}&=&{\cal P}_{i}(r)= (r^{2}+a^{2})E_{i}-aL_{i},
\end{eqnarray} 
and $i=1,2$.
Thus, the CM energy can be given in terms of $\mu_{i}$, $E_{i}$, $L_{i}$, 
and $r$.
If we assume that $\sigma_{1r}$ and $\sigma_{2r}$ are of the same sign,
Eq.~(\ref{eq:E_cm_explicit_equatorial}) for the near-horizon limit then reduces to
\begin{eqnarray}
E_{\rm cm}^{2}&=&\mu_{1}^{2}+\mu_{2}^{2}+\frac{1}{r_{H}^{2}}
\left\{\left[\mu_{1}^{2}r_{H}^{2}+(L_{1}-aE_{1})^{2}\right]\frac{E_{2}-\Omega_{H}L_{2}}{E_{1}-\Omega_{H}L_{1}}\right.\nonumber \\
&& \left. +\left[\mu_{2}^{2}r_{H}^{2}+(L_{2}-aE_{2})^{2}\right]\frac{E_{1}-\Omega_{H}L_{1}}{E_{2}-\Omega_{H}L_{2}}-2(L_{1}-a E_{1})(L_{2}-a E_{2})\right\}.
\label{eq:Harada_Kimura_variant}
\end{eqnarray}
It is clear that the necessary condition for the CM energy to be arbitrarily 
high is that $(E-\Omega_{H} L)$ is arbitrarily close to zero for either of the 
two particles.

\section{Transition from adiabatic inspiral to plunge}
We here briefly review the formalism of the transition from 
adiabatic inspiral to plunge 
proposed by Ori and Thorne~\cite{Ori:2000zn} 
and extended by Kesden~\cite{Kesden:2011ma}.

\subsection{Ori-Thorne formalism}
The geodesic equation and the normalisation of the four-velocity 
of a massive particle of rest mass $\mu$ 
in the Kerr spacetime are given by 
\begin{equation}
\frac{d^{2}\tilde{r}}{d\tilde{\tau}^{2}}
=-\frac{1}{2}\frac{\partial V}{\partial \tilde{r}},
\label{eq:geodesic}
\end{equation}
and 
\begin{equation}
\left(\frac{d\tilde{r}}{d\tilde{\tau}}\right)^{2}
=\tilde{E}^{2}-V,
\label{eq:normalisation}
\end{equation}
respectively, 
where the effective potential $V=V(\tilde{r},\tilde{E},\tilde{L})$ is given by
\begin{equation}
V(\tilde{r},\tilde{E},\tilde{L})=1-\frac{2}{\tilde{r}}
+\frac{\tilde{L}^{2}+\tilde{a}^{2}-\tilde{E}^{2}\tilde{a}^{2}}{\tilde{r}^{2}}-\frac{2(\tilde{L}-\tilde{E}\tilde{a})^{2}}{\tilde{r}^{3}}
\label{eq:effective_potential}
\end{equation}
and we define nondimensional quantities $\tilde{r}=r/M$, 
$\tilde{t}=t/M$, $\tilde{a}=a/M$, $\tilde{\tau}=\tau/M$,
$\tilde{E}=E/\mu$, and $\tilde{L}=L/(\mu M)$.

We first expand the effective potential in the Taylor
series around the ISCO radius, energy, and angular momentum, i.e.
$(\tilde{r},\tilde{E},\tilde{L})=
(\tilde{r}_{\rm ISCO},\tilde{E}_{\rm ISCO},\tilde{L}_{\rm ISCO})$,  
in terms of $R=\tilde{r}-\tilde{r}_{\rm ISCO}$,
$\chi=\tilde{\Omega}^{-1}_{\rm ISCO}(\tilde{E}-\tilde{E}_{\rm ISCO})$, 
and 
$\xi=\tilde{L}-\tilde{L}_{\rm ISCO}$ up to $O(R^{3},\chi,\xi)$,
where $\Omega=(d\phi/dt)=\tilde{\Omega}/M$ is the angular velocity of a particle 
 and $\tilde{\Omega}_{\rm ISCO}$ is $\tilde{\Omega}$ for a particle orbiting 
the ISCO.
Then, the geodesic equation (\ref{eq:geodesic}) becomes
\begin{equation}
\frac{d^{2}R}{d\tilde{\tau}^{2}}=-\alpha R^{2}+\beta\xi-
\frac{1}{2}\left(\tilde{\Omega}\frac{\partial^{2}V}
{\partial\tilde{E}\partial\tilde{r}}\right)_{\rm ISCO}(\chi-\xi)+\cdots,
\label{eq:geodesic_Taylor}
\end{equation}
where 
\begin{eqnarray}
\alpha=\frac{1}{4}\left(\frac{\partial ^{3}V}{\partial \tilde{r}^{3}}\right)_{\rm ISCO}, \quad 
\beta = -\frac{1}{2}\left(\frac{\partial^{2} V}{\partial \tilde{L} \partial \tilde{r}}+\tilde{\Omega}\frac{\partial^{2} V}{\partial {\tilde{E}} \partial \tilde{r}}\right)_{\rm ISCO},
\end{eqnarray}
and the subscript ISCO means the value estimated 
at $(\tilde{r},\tilde{E},\tilde{L})=(\tilde{r}_{\rm ISCO}, \tilde{E}_{\rm ISCO},\tilde{L}_{\rm ISCO})$.

To take radiation reaction into account,
we introduce $\kappa$ as follows:
\begin{equation}
\kappa\equiv -\left(\tilde{\Omega}^{-1}\eta^{-2}\frac{dE}{dt}\frac{d\tilde{t}}{d\tilde{\tau}}\right)_{\rm ISCO}=-\left(\tilde{\Omega}^{-1}\eta^{-1}\frac{d\tilde{E}}{d\tilde{t}}
\frac{d\tilde{t}}{d\tilde{\tau}}\right)_{\rm ISCO},
\end{equation}
where $\eta\equiv \mu/M $ is the mass ratio. 
It is $\kappa$ that drives the 
object in the radial direction. 
We assume that the loss of the object's energy 
is radiated away through gravitational radiation, i.e.
\begin{equation}
-\left(\frac{dE}{dt}\right)=\dot{E}_{\rm GW}=\frac{32}{5}\eta^{2}\tilde{\Omega}^{10/3}\dot{\mathcal E},
\label{eq:energy_balance}
\end{equation}
where $\dot{\mathcal E}$ is the nondimensional 
correction factor to the Newtonian 
quadrupole formula for the gravitational wave luminosity~\cite{Finn:2000sy}. $\dot{\mathcal E}$ contains all the 
relativistic effects including those from the spin of the black hole.
$\kappa$ is then rewritten as
\begin{equation}
\kappa=\frac{32}{5}\left(\tilde{\Omega}^{7/3}\frac{d\tilde{t}}{d\tilde{\tau}}
\dot{\mathcal E}\right)_{\rm ISCO}.
\label{eq:kappa}
\end{equation}

For the neighboring circular orbits, we find 
$\delta E=\Omega \delta L$ (see e.g. Ref.~\cite{Page:1974he}).
It suggests $\chi=\xi$ and  
we can take radiation reaction into account 
from Eq.~(\ref{eq:energy_balance}),
\begin{equation}
\chi=\xi=-\eta\kappa\tilde{\tau}.
\end{equation} 
In terms of the redefined variables, 
Eq.~(\ref{eq:geodesic_Taylor}) yields
\begin{equation}
\ddot{X}=-X^{2}-T,
\label{eq:OT_1}
\end{equation}
where 
\begin{eqnarray}
R=\eta^{2/5}R_{0}X,\quad
\tilde{\tau}=\eta^{-1/5}\tau_{0}T, \quad
R_{0}= (\beta\kappa)^{2/5}\alpha^{-3/5}, \quad
\tau_{0}= (\alpha\beta\kappa)^{-1/5},
\end{eqnarray}
and the dot in Eq.~(\ref{eq:OT_1}) denotes the derivative with respect to $T$.

Ori and Thorne numerically obtained a unique solution 
to Eq.~(\ref{eq:OT_1}) with the initial condition 
\begin{equation}
X\approx \sqrt{-T}
\label{eq:early-time_behaviour}
\end{equation}
as $T\to -\infty$. With this condition it is assumed that the 
object orbits circularly at the potential minimum, which 
moves inwardly adiabatically at early times.
The solution is shown in Figs. 2 and 3 of Ref.~\cite{Ori:2000zn}.
This solution monotonically decreases with $T$ and 
diverges to negative infinity at $T=T_{\rm div}$~\footnote{We here 
denote the time of divergence as $T_{\rm div}$
rather than $T_{\rm plunge}$.} as
\begin{equation}
X\approx -\frac{6}{(T_{\rm div}-T)^{2}},
\label{eq:X_plunge}
\end{equation}
where $T_{\rm div}\simeq 3.412$~\cite{Ori:2000zn,Kesden:2011ma}. 
This divergence should be 
regarded as the breakdown of the Taylor-series expansion at
very large values of $|X|$.

The location of the ISCO, $\tilde{r}=\tilde{r}_{\rm ISCO}$, 
in terms of $X$ is of course given by $X=0$.
The location of the horizon $\tilde{r}=\tilde{r}_{H}$
in terms of $X$ is given by  
\begin{equation}
X_{H}=\eta^{-2/5}\frac{\tilde{r}_{H}-\tilde{r}_{\rm ISCO}}{R_{0}}.
\end{equation}
We denote the time $T$ when the object crosses the ISCO radius as $T_{0}$, i.e. $X(T_{0})=0$, while the time $T$ when it crosses the event horizon
as $T_{H}$, i.e. $X(T_{H})=X_{H}$. The numerical value of $T_{0}$ is given by 
$T_{0}\simeq 0.72$~\cite{Kesden:2011ma}. Clearly, $T_{0}<T_{H}<T_{\rm div}$
holds.

\subsection{Kesden's extension}
Through the Taylor-series expansion around the ISCO particle, 
the normalisation  (\ref{eq:normalisation}) becomes
\begin{equation}
\left(\frac{dR}{d\tilde{\tau}}\right)^{2}=-\frac{2\alpha}{3}R^{3}+2\beta R\xi
{
+\left(\frac{\partial V}{\partial \tilde{L}}\right)_{\rm ISCO}(\chi-\xi)-}
\left(\tilde{\Omega}\frac{\partial^{2} V}{\partial \tilde{E}\partial \tilde{r}}\right)_{\rm ISCO}(\chi-\xi)R+\cdots.
\label{eq:normalisation_Taylor}
\end{equation}
Taking the same procedure as in the derivation of Eq.~(\ref{eq:OT_1}), we reduce Eq.~(\ref{eq:normalisation_Taylor}) to
\begin{equation}
\dot{X}^{2}=-\frac{2}{3}X^{3}-2XT.
\label{eq:Ori-Thorne}
\end{equation}
It can be seen that 
the Ori-Thorne solution does not satisfy 
Eq.~(\ref{eq:Ori-Thorne}).
Noting that the relation $\chi=\xi$ is required 
only for the quasicircular orbits, Kesden~\cite{Kesden:2011ma} introduces
$Y$ in Eq.~(\ref{eq:normalisation_Taylor}) through
\begin{equation}
\chi-\xi=\eta^{6/5}(\chi-\xi)_{0}Y,\quad 
(\chi-\xi)_{0}=
\alpha^{-4/5}(\beta\kappa)^{6/5}\left(\frac{\partial V}{\partial \tilde{L}}\right)^{-1}_{\rm ISCO}.
\label{eq:chi-xi}
\end{equation}
Then, we obtain 
\begin{equation}
\dot{X}^{2}=-\frac{2}{3}X^{3}-2XT+Y.
\label{eq:normalisation_Y}
\end{equation}
To restore the consistency between the 
equation of motion~(\ref{eq:OT_1}) and
the normalisation relation (\ref{eq:normalisation_Y}), 
$Y$ must satisfy 
\begin{equation}
\dot{Y}=2X.
\label{eq:OT_2}
\end{equation}

The solution for $Y$ to Eq.~(\ref{eq:OT_2}) is 
numerically obtained and shown in Fig. 3 of Ref.~\cite{Kesden:2011ma}. In particular, the solution shows the asymptotic behaviours
\begin{equation}
Y\approx -\frac{4}{3}(-T)^{3/2}
\end{equation}
for $T\to -\infty$, and
\begin{equation}
Y\approx -\frac{12}{T_{\rm div}-T},
\label{eq:Y_plunge}
\end{equation}
for $T\to T_{\rm div}$. 

The energy or angular momentum of the object is now not conserved 
because the object no longer moves along a geodesic of the background 
geometry. The energy and angular momentum change as
\begin{eqnarray}
\tilde{E}=\tilde{E}_{\rm ISCO}+\Delta \tilde{E}_{\rm tr}+\Delta \tilde{E}_{\rm norm}, \quad 
\tilde{L}=\tilde{L}_{\rm ISCO}+\Delta \tilde{L}_{\rm tr},
\end{eqnarray}
where 
\begin{eqnarray}
\Delta \tilde{E}_{\rm tr}=\tilde{\Omega}_{\rm ISCO} \Delta \tilde{L}_{\rm tr}=
-\tilde{\Omega}_{\rm ISCO}\eta^{4/5}\kappa \tau_{0}T, \quad
\Delta\tilde{E}_{\rm norm}=\tilde{\Omega}_{\rm ISCO}\eta^{6/5}(\chi-\xi)_{0}Y.
\end{eqnarray}
It should be noted that the correction to restore the normalisation of the four-velocity may also be added to the angular momentum of the object. This ambiguity does not affect our conclusion 
in the present paper.

It is natural to take Eq.~(\ref{eq:chi-xi}) into account also in
Eq.~(\ref{eq:geodesic_Taylor}). 
This implies
\begin{equation}
\ddot{X}=-X^{2}-T+\epsilon Y,
\label{eq:xddot_kesden}
\end{equation}
where 
\begin{eqnarray*}
\epsilon=\eta^{2/5}C, \quad 
C=-\frac{1}{2}\alpha^{-3/5}(\beta\kappa)^{2/5}
\left[\tilde{\Omega} 
\frac{\partial^{2} V}{\partial \tilde{E}\partial \tilde{r}}\left(\frac{\partial V}{\partial \tilde{L}}\right)^{-1}\right]_{\rm ISCO}.
\end{eqnarray*}
Thus, Eq.~(\ref{eq:xddot_kesden}) justifies the Ori-Thorne solution as 
a solution if $\epsilon \ll 1$.

The consistency of Eq.~(\ref{eq:xddot_kesden})
with the normalisation relation (\ref{eq:normalisation_Y}) 
of the four-velocity requires that $Y$ must satisfy
\begin{equation}
\dot{Y}=2X+2\epsilon Y\dot{X}.
\label{eq:ydot_kesden}
\end{equation}
Equivalently, we can eliminate $Y$ from 
Eq.~(\ref{eq:normalisation_Y}) and obtain from Eq.~(\ref{eq:xddot_kesden})
\begin{equation}
\ddot{X}=-X^{2}-T+\epsilon\left(\dot{X}^{2}+\frac{2}{3}X^{3}+2XT\right).
\end{equation}
Kesden numerically integrated Eqs.~(\ref{eq:xddot_kesden}) and (\ref{eq:ydot_kesden}) simultaneously with different values of $\epsilon$ and 
the solutions are shown in Fig.~6 of Ref.~\cite{Kesden:2011ma}.
As for the initial values for $X$, $\dot{X}$, and $Y$, 
$\dot{X}=\ddot{X}=0$ is assumed in Eqs.~(\ref{eq:normalisation_Y}) and (\ref{eq:xddot_kesden}) and $X$ and $Y$ are solved algebraically at some small value of $T$~\cite{Kesden:private}.
Since Eq.~(\ref{eq:early-time_behaviour}) no longer provides a proper asymptotic solution of Eqs.~(\ref{eq:xddot_kesden}) and (\ref{eq:ydot_kesden}) for $\epsilon\ne 0$, this choice of the initial condition is one of the natural 
choices as the matching to the adiabatic inspiral phase at early times.
We can see in Fig.~6 of Ref.~\cite{Kesden:2011ma} that the numerical solutions 
are very close to the Ori-Thorne solution  ($\epsilon=0$) 
for $\epsilon\ll 1$ and behave qualitatively similarly 
even for $\epsilon\sim 1$.
Although one can still obtain numerical solutions 
to Eqs.~(\ref{eq:xddot_kesden}) and ~(\ref{eq:ydot_kesden}) with 
this initial condition even if $\epsilon\agt 1$, 
such numerical solutions may probably be invalid because 
higher-order terms in the Taylor-series expansions 
or higher-order terms in $\epsilon$
should not be negligible.
We can here only assume that the Ori-Thorne-Kesden formalism 
is justified so that 
the numerical solutions for $X$ and $Y$
for $\epsilon=0$, i.e., the Ori-Thorne
solution to Eq.~(\ref{eq:OT_1})
for $X$ together with Kesden's solution to Eq.~(\ref{eq:OT_2})
for $Y$, qualitatively give the right
behaviours for $\epsilon\alt 1$.
Hereafter, we
restrict the analysis within the regime
$\epsilon\alt 1$.

\section{maximal rotation limit}
In the maximal rotation limit
$\delta=1-\tilde{a}\to 0$, 
the numerical results by the GREMLIN code can be fit by 
\begin{equation}
\dot{\mathcal E}=A\delta^{m},
\label{eq:GREMLIN} 
\end{equation}
where $A\simeq 1.80$ and 
$m\simeq 0.317$~\cite{Kesden:2011ma}.
There is another argument by Chrzanowski~\cite{Chrzanowski:1976jy}
which suggests $m\simeq 1/3$~\cite{Kesden:2011ma}.  

We can obtain the dependence of the quantities 
on $\delta=1-\tilde{a}\ll 1$ as follows:
\begin{eqnarray}
\tilde{r}_{H}\simeq  1+(2\delta)^{1/2}, \quad
\tilde{\Omega}_{H}\simeq \frac{1}{2}\left[1-(2\delta)^{1/2}\right]
\end{eqnarray}
for the black hole event horizon, 
\begin{eqnarray}
\tilde{r}_{\rm ISCO}&\simeq & 1+(4\delta)^{1/3}+\frac{7}{8}(4\delta)^{2/3}, \\
\tilde{\Omega}_{\rm ISCO}&\simeq &\frac{1}{2}\left[1-\frac{3}{4}(4\delta)^{1/3}-\frac{9}{32}(4\delta)^{2/3}\right], 
\label{eq:Omega_ISCO}\\
\tilde{E}_{\rm ISCO}&\simeq & \frac{1}{\sqrt{3}}\left[1+(4\delta)^{1/3}-\frac{5}{8}(4\delta)^{2/3}\right], \\
\tilde{L}_{\rm ISCO}&\simeq& \frac{2}{\sqrt{3}}\left[1+(4\delta)^{1/3}+\frac{1}{8}(4\delta)^{2/3}\right], \\
\left(\frac{d\tilde{t}}{d\tilde{\tau}}\right)_{\rm ISCO}
&\simeq & \frac{4}{\sqrt{3}}(4\delta)^{-1/3}\left[1-\frac{3}{8}(4\delta)^{1/3}+\frac{7}{32}(4\delta)^{2/3}\right]
\label{eq:dtdtau_ISCO}
\end{eqnarray}
for the ISCO particle~\cite{Bardeen:1972fi,Harada:2010yv},
and 
\begin{eqnarray}
\alpha &\simeq & 1-4(4\delta)^{1/3}, \\
\beta&\simeq & \frac{\sqrt{3}}{2}(4\delta)^{1/3}, \\
\left(\frac{\partial V}{\partial \tilde{L}}\right)_{\rm ISCO} &\simeq & \frac{4}{\sqrt{3}}(4\delta)^{1/3}, \\
\left(\frac{\partial^{2} V}{\partial \tilde{E}\partial \tilde{r}}
\right)_{\rm ISCO}&\simeq &
-\frac{8}{\sqrt{3}}\left[1-\frac{7}{2}(4\delta)^{1/3}\right]
\end{eqnarray}
from Eq.~(\ref{eq:effective_potential})
for the derivatives of the effective potential 
for the ISCO particle. 
As for the dynamics driven by gravitational radiation, 
from Eqs.~(\ref{eq:kappa}), (\ref{eq:GREMLIN}), (\ref{eq:Omega_ISCO}),
and (\ref{eq:dtdtau_ISCO}), 
we obtain
\begin{eqnarray}
\kappa\simeq  \frac{16}{5\sqrt{3}}A\delta ^{m-1/3},
\end{eqnarray}
and then $R_{0}$, $\tau_{0}$, $(\chi-\xi)_{0}$, $C$, and $\epsilon$ 
are written in terms of $\delta $ and $\eta$
as follows:
\begin{eqnarray}
R_{0}&\simeq  & 2^{22/15}5^{-2/5}A^{2/5}\delta^{2m/5}, \\
\tau_{0}&\simeq  & 2^{-11/15}5^{1/5}A^{-1/5}\delta^{-m/5}, \\
(\chi-\xi)_{0}&\simeq & 2^{26/15}3^{1/2}5^{-6/5}A^{6/5}
\delta^{6m/5-1/3}, \\
C&\simeq &2^{-1/5}5^{-2/5}A^{2/5}\delta^{2m/5-1/3}, \\
\epsilon&\simeq & 2^{-1/5}5^{-2/5}A^{2/5}\eta^{2/5}\delta^{2m/5-1/3}.
\label{eq:epsilon_near-extremal}
\end{eqnarray}

Since $C$ is divergent as $\delta\to 0$ if $m<5/6$, 
the parameter $\epsilon$ is divergent if we take the maximal rotation limit 
$\delta\to 0$ as the mass ratio $\eta$ is kept constant.
In this case, the 
Ori-Thorne-Kesden formalism may be invalid. 
Clearly, the two limits $\delta\to 0$ and $\eta\to 0$ cannot be taken
independently.
As $\delta$ is kept constant, 
we can always take the limit $\eta\to 0$, where
the Ori-Thorne-Kesden formalism is justified.
Note that Eq.~(\ref{eq:epsilon_near-extremal}) 
can be solved for $\eta$ as follows:
\begin{equation}
\eta\simeq 5\sqrt{2} A^{-1}\delta ^{5/6-m}\epsilon^{5/2}.
\end{equation}
This implies that the present transition 
formalism is valid for a rather wide range of
the mass ratio if we consider a reasonable value of 
the black hole spin.

The time varying parts of the 
energy and angular momentum of the object in the transition are given by
\begin{eqnarray}
\Delta \tilde{E}_{\rm tr}&=&\tilde{\Omega}_{\rm ISCO}\Delta \tilde{L}_{\rm tr}\simeq-2^{34/15}3^{-1/2}5^{-4/5}A^{4/5}\eta^{4/5}\delta^{4m/5-1/3}T \nonumber  \\
&\simeq& -2^{8/3}3^{-1/2}\delta^{1/3}\epsilon^{2}T, \\
\Delta\tilde{E}_{\rm norm}&\simeq&2^{11/15}3^{1/2}5^{-6/5}A^{6/5}\eta^{6/5}\delta ^{6m/5-1/3} Y \nonumber \\
&\simeq& 2^{4/3}3^{1/2}\delta^{2/3} \epsilon^{3}Y.
\end{eqnarray}
If we substituted $T=T_{\rm div}$, $Y$ would diverge to negative infinity and hence $\Delta \tilde{E}_{\rm norm}$ would diverge. 
However, since we are interested in the CM energy of two objects colliding outside the event horizon, we should stop the calculation at $T=T_{H}$. Noting
\begin{equation}
X_{H}\simeq  -2^{-4/5}5^{2/5}A^{-2/5}\eta^{-2/5}\delta^{-2m/5+1/3}
\simeq -\frac{1}{2\epsilon},
\end{equation}
we find $T_{\rm div}-T_{H}\approx 2\sqrt{3\epsilon}$ from Eq.~(\ref{eq:X_plunge}). Hence, Eq.~(\ref{eq:Y_plunge}) implies $Y(T_{H})\simeq -2\sqrt{3}\epsilon^{-1/2}$ and then 
\begin{eqnarray}
\Delta \tilde{E}_{\rm tr}&\simeq & 
-2^{8/3}3^{-1/2}\delta^{1/3}\epsilon^{2}T_{H}, \\
\Delta \tilde{E}_{\rm norm}&\simeq & -2^{7/3}3\delta^{2/3}\epsilon^{5/2}.
\label{eq:E_norm_TH}
\end{eqnarray}
Therefore, the energy and angular momentum 
extracted 
through gravitational waves should be finite
until the object plunges into the horizon.  

\section{CM energy for the collision of an object in the transition}

The CM energy can be directly calculated in terms of the four-velocities of the two colliding objects. Since the four-velocity can be uniquely expressed by $\tilde{r}$, $\tilde{E}$, and $\tilde{L}$ for the equatorial motion 
in the Kerr spacetime, we can use the formula for the CM energy in terms of $\tilde{E}$ and $\tilde{L}$ of each particle using their values at the moment of collision. The formula (\ref{eq:Harada_Kimura_variant}) implies that the CM energy can be arbitrarily high if the quantity $(E-\Omega_{H}L)$ is arbitrarily close to zero at the moment of collision. If we consider radiation reaction, this quantity is no longer conserved. 

To examine  whether the CM energy is bounded or not, we only have to see
whether the quantity $(E-\Omega_{H}L)$ is vanishing or not in the limit to 
the event horizon. We can rewrite $(E-\Omega_{H}L)$ as follows:
\begin{eqnarray}
E-\Omega_{H}L&=& E-E_{\rm ISCO}-\Omega_{\rm ISCO}(L-L_{\rm ISCO})
-(\Omega_{H}-\Omega_{\rm ISCO})(L-L_{\rm ISCO})
+(E_{\rm ISCO}-\Omega_{H}L_{\rm ISCO}) \nonumber \\
&=& \mu\left[\Delta \tilde{E}_{\rm norm}-
(\tilde{\Omega}_{H}-\tilde{\Omega}_{\rm ISCO})\Delta \tilde{L}_{\rm tr}
+(\tilde{E}_{\rm ISCO}-\tilde{\Omega}_{H}\tilde{L}_{\rm ISCO})\right],
\end{eqnarray}
where we can find
\begin{eqnarray}
-(\tilde{\Omega}_{H}-\tilde{\Omega}_{\rm ISCO})\Delta \tilde{L}_{\rm tr}
&\simeq& 2^{14/15}3^{1/2}5^{-4/5}A^{4/5}\eta^{4/5}\delta^{4m/5}T 
\label{eq:contribution2} \nonumber \\
&\simeq & 2^{4/3}3^{1/2}\epsilon^{2}\delta ^{2/3}T, \\
\tilde{E}_{\rm ISCO}-\tilde{\Omega}_{H}\tilde{L}_{\rm ISCO}&\simeq & \frac{1}{\sqrt{3}}(2\delta)^{1/2}.
\label{eq:contribution3}
\end{eqnarray}

We can now compare Eqs.~(\ref{eq:E_norm_TH}), (\ref{eq:contribution2}), and (\ref{eq:contribution3}) at $T=T_{H}$.
In the limit $\delta\to 0$ as $\epsilon$ is kept constant, we find
that $(\tilde{E}_{\rm ISCO}-\tilde{\Omega}_{H}\tilde{L}_{\rm ISCO})$ 
gives a dominant contribution. Therefore, the 
formula for the 
CM energy for the collision of an ISCO particle with a generic particle
obtained in Ref.~\cite{Harada:2011xz,Harada:2010yv} is applicable.
The result is the following:
\begin{equation}
E_{\rm cm}\simeq \frac{2^{1/4}}{3^{1/4}}\sqrt{\mu_{1}\mu_{2}}\frac{\sqrt{2\tilde{E_{2}}-\tilde{L_{2}}}}{\delta^{1/4}},
\end{equation}
where we assume object 1 is in the transition while 
object 2 is a generic counterpart.
Because of the condition $\epsilon\alt 1$,  
there appears a maximum value of the CM energy, 
which weakly depends on the mass 
ratio $\eta_{1}$:
\begin{equation}
E_{\rm cm}\simeq \left(\frac{2}{3}\right)^{1/4}\left(\frac{A}{5\sqrt{2}}\right)^{-3/[2(5-6m)]}\sqrt{2\tilde{E}_{2}-\tilde{L}_{2}} \sqrt{M\mu_{2}}
\eta_{1}^{(1-3m)/(5-6m)}\epsilon^{15/[4(5-6m)]}.
\end{equation}
Assuming $m=1/3$, $A=1.8$, $\tilde{E}_{2}=1$, and $\tilde{L}_{2}=0$, we find 
\begin{eqnarray}
E_{\rm cm}
&\simeq& 2.6\times 10^{30} \mbox{GeV} \left(\frac{\mu_{2}}{100 \mbox{GeV}}\right)^{1/2}
\left(\frac{M}{10 M_{\odot}}\right)^{1/2}\epsilon^{5/4} \\
&\simeq & 4.6\times 10^{58} \mbox{erg} \left(\frac{\mu_{2}}{M_{\odot}}\right)^{1/2} \left(\frac{M}{10^{8}M_{\odot}}\right)^{1/2}\epsilon^{5/4},
\end{eqnarray}
and hence the maximum value realised for $\epsilon\simeq 1$
will not depend on the mass ratio of 
the object in the transition. If the object in the transition collides with 
a dark matter particle of mass 100 GeV around a stellar mass 
black hole, the CM energy can be much greater than the Planck energy.
If the collision counterpart 
is a stellar mass compact object around a supermassive black hole,
the CM energy can be as energetic as $10^{58}$ erg.

It is also interesting to see the CM energy 
for the collision at the ISCO radius. 
In this case, we cannot take the near-horizon limit before taking 
the maximal rotation limit. If we neglect 
radiation reaction,
the particle orbiting the ISCO has a vanishing radial velocity 
by construction and we obtain~\cite{Harada:2011xz,Harada:2010yv}
\begin{equation}
E_{\rm cm}\simeq \frac{2^{2/3}}{3^{1/4}}\sqrt{\mu_{1}\mu_{2}}\frac{\sqrt{2\tilde{E_{2}}-\tilde{L_{2}}}}{\delta^{1/6}}.
\label{eq:ON-ISCO_formula}
\end{equation}
When we take radiation reaction into account, since the object in the transition 
has a nonvanishing radial velocity at the ISCO radius, it is not trivial 
whether or not the expression given by Eq.~(\ref{eq:ON-ISCO_formula}) is still valid.
Indeed, substituting the expressions 
for $\tilde{E}$ and $\tilde{L}$ obtained in the previous 
section into the general equatorial formula (\ref{eq:E_cm_explicit_equatorial}) 
and evaluating it at the ISCO radius, 
we can find that the above expression gives the leading order term
in the limit $\delta\to 0$ as $\epsilon$ is kept constant.
Then, the condition $\epsilon\alt 1$ taken into account, 
we estimate $E_{\rm cm}$ as follows: 
\begin{equation}
E_{\rm cm}\simeq 2^{2/3}3^{-1/4}\left(\frac{A}{5\sqrt{2}}\right)^{-1/(5-6m)}
\sqrt{2\tilde{E}_{2}-\tilde{L}_{2}}
\mu_{1}^{3(1-2m)/[2(5-6m)]}\mu_{2}^{1/2}M^{1/(5-6m)}
\epsilon^{5/[2(5-6m)]}.
\end{equation}
Assuming $m=1/3$, $A=1.8$, $\tilde{E}_{2}=1$, and $\tilde{L}_{2}=0$, 
we find 
\begin{eqnarray}
E_{\rm cm}
&\simeq& 1.3\times 10^{21} \mbox{GeV} \left(\frac{\mu_{1}}{100 \mbox{GeV}}\right)^{1/6}
\left(\frac{\mu_{2}}{100 \mbox{GeV}}\right)^{1/2}
\left(\frac{M}{10 M_{\odot}}\right)^{1/3}\epsilon^{5/6} \\
&\simeq & 2.3\times 10^{57} \mbox{erg} \left(\frac{\mu_{1}}{M_{\odot}}\right)^{1/6}
\left(\frac{\mu_{2}}{M_{\odot}}\right)^{1/2}
\left(\frac{M}{10^{8}M_{\odot}}\right)^{1/3}
\epsilon^{5/6}.
\end{eqnarray}
Thus the CM energy is lower than that for the near-horizon collision
but still significantly high.

It should be noted that 
the fact that the maximum CM energy that can be reached 
within the present framework is extremely high suggests
that the collision with reasonably high CM energy occurs rather 
frequently, although the precise estimate of its frequency 
is out of the scope of the present paper.

\section{Conclusions}

When we consider the collision of two colliding particles around 
a nearly maximally rotating Kerr black hole, the CM energy of the particles
can be arbitrarily high if gravitational radiation is neglected.
Although the originally proposed scenario through direct collision
from infinity 
needs an artificial fine-tuning of the angular momentum of either of the 
particles, it turns out that the fine-tuning is naturally realised 
for a particle orbiting the ISCO in the EMRI.
We have studied this scenario with gravitational 
radiation reaction,
where the object experiences a continuous transition
from adiabatic inspiral to plunge into the horizon.
Applying the Ori-Thorne-Kesden formalism of transition, 
we have found that it is 
gravitational radiation reaction that realises the 
fine-tuning of the angular momentum and the 
expression for the CM energy is not affected.
Then, we have discussed
how high the CM energy can reach within the condition where the 
present transition formalism is well justified.
We find that the CM energy can still be high enough to be 
of great physical interest.
However, it should be noted that the present analysis incorporates
some but not all the effects of self-force. A systematic 
approach is necessary to study the effects of conservative
self-force on the problem of two-body collision around a rapidly rotating black hole. 

Finally, we comment on the possibilities and difficulties of observing the consequences of the high-velocity collisions. It is shown in Ref.~\cite{Jacobson:2009zg} that the Killing energy of the ejecta particle from the high-energy collision of two particles of rest mass $m$ is at most $2m$. This can be explained by the effect of strong redshift. Thus, it is not expected that high-energy ejecta particles can be directly observed by a distant observer. On the other hand, it is suggested in e.g. Refs.~\cite{Banados:2009pr,Harada:2011xz,Banados:2010kn,Williams:2011uz,Harada:2010yv} 
that some indirect signatures of the high-energy collision of particles near the black hole horizon might be observed by means of electromagnetic waves and/or gravitational waves. Further studies are necessary to reveal what observational signatures are expected.

\acknowledgments
The authors thank E.~Barausse, T.~Igata, S.~Isoyama, S. Jhingan, M.~Kesden, U.~Miyamoto, K.~Nakao, J.~Novak, N.~Sago, M.~Saijo, N.~Shibazaki, R.~Takahashi, and T.~Tanaka
for very helpful comments and suggestions. 
T.H. was supported by a Grant-in-Aid for Scientific Research from the Ministry of Education, Culture, Sports, Science, and Technology of Japan [Young Scientists (B) No. 21740190].
M.K. was supported by the JSPS Grant-in-Aid for Scientific Research No.23$\cdot$2182.

\appendix*

\section{Final spin of the black hole in the merger}
The present estimate also applies to the final spin 
after a rapidly 
spinning black hole swallows an inspiralling 
object. One can estimate the final spin of the black hole 
in the merger as follows:
\begin{eqnarray}
\tilde{a}_{f}&=&\frac{\tilde{a}+\eta (\tilde{L}+\Delta \tilde{L}_{\rm tr})}{[1+\eta (\tilde{E}_{\rm ISCO}+\Delta E_{\rm tr}+\Delta E_{\rm norm})]^{2}}  \nonumber \\
&=& \tilde{a}+\eta(\Delta \tilde{a}_{\rm ISCO}+\Delta \tilde{a}_{\rm tr}+\Delta \tilde{a}_{\rm norm}),
\end{eqnarray}
where 
\begin{eqnarray}
\Delta \tilde{a}_{\rm ISCO}&=&\tilde{L}_{\rm ISCO}-2 \tilde{E}_{\rm ISCO}\propto \delta^{2/3}, 
\label{eq:a_ISCO}\\
\Delta \tilde{a}_{\rm tr}&=&\Delta \tilde{L}_{\rm tr}-2\Delta \tilde{E}_{tr}\propto -\delta^{2/3}\epsilon^{2}T_{H} \propto -\delta^{2/3}\epsilon^{2}, 
\label{eq:a_tr}\\
\Delta \tilde{a}_{\rm norm}&=&-2\Delta \tilde{E}_{\rm norm}\propto -\delta^{2/3}\epsilon^{3}Y_{H}\propto \delta^{2/3}\epsilon^{5/2}.
\label{eq:a_norm}
\end{eqnarray}
Note that we should take $T=T_{H}$ for this estimate. We can see that 
$\Delta \tilde{a}_{\rm tr}$ and $\Delta\tilde{a}_{\rm norm}$
cannot dominate $\Delta \tilde{a}_{\rm ISCO}$ so that  
the effects of radiation reaction do not 
prevent or promote the 
overspinning of the black hole.
The above estimate is slightly different 
from that in Ref.~\cite{Kesden:2011ma}, 
where $\Delta \tilde{a}_{\rm norm}$ is estimated to be
{\it negative} and {\it dominant} 
if $\delta\to 0$ as $\eta (>0)$ is kept constant. 
In the present analysis, since we assume
$\epsilon\alt 1$, we 
obtain Eqs.~(\ref{eq:a_ISCO})-(\ref{eq:a_norm}), where 
all three are well controlled. If we further 
assume $\epsilon\ll 1$, we can see that $\Delta \tilde{a}_{\rm ISCO}$
is positive and dominant, 
$\Delta \tilde{a}_{\rm tr}$ is negative and subdominant and 
$\Delta \tilde{a}_{\rm norm}$ is {\it positive} and {\it subdominant}.
Therefore, the effects of radiation reaction are subdominant
within the transition formalism and this is consistent with 
the result in Ref.~\cite{Barausse:2010ka}.
Although we will still need 
to consider the contribution of 
ingoing gravitational waves into the horizon, 
this would not change our conclusion.
Thus, it is suggested that for $\epsilon\ll 1$,
radiation reaction would not play an important role in 
an attempt to overspin a black hole by plunging an object
and the conservative part of the self-force 
should be critical, 
which we have neglected in the present analysis.

\end{document}